МІНІСТЕРСТВО ОСВІТИ І НАУКИ УКРАЇНИ

ДОНЕЦЬКИЙ НАЦІОНАЛЬНИЙ УНІВЕРСИТЕТ ІМЕНІ ВАСИЛЯ СТУСА

**СОЛОГУБ ВЛАДИСЛАВ АНДРІЙОВИЧ**

Допускається до захисту:
Завідувач кафедри
інформаційних технологій,
доктор технічних наук, доцент
___________Т. В. Нескородєва
__________________________
«____»______________2022 р.

# РОЗРОБКА ІНФОРМАЦІЙНОЇ СИСТЕМИ ДЛЯ СТАТИСТИЧНОГО АНАЛІЗУ РОЗПОДІЛІВ ГЛОБАЛЬНИХ БРЕНДІВ

Спеціальність 122 «Комп'ютерні науки»

**Кваліфікаційна (бакалаврська) робота**

Керівник:
Штовба С. Д., д.т.н., професор,
професор кафедри інформаційних технологій

Оцінка: ___/___/_______
(бали за шкалою СКТС/за національною шкалою)

Голова ЕК: ___________
(підпис)

Вінниця - 2022

# АНОТАЦІЯ

**Сологуб В. А. Розробка інформаційної системи для статистичного аналізу розподілу глобальних брендів.** Спеціальність 122 «Комп’ютерні науки». Донецький національний університет імені Василя Стуса, Вінниця, 2022.

У кваліфікаційній (бакалаврській) роботі досліджено методи статистичного аналізу розподілів глобальних брендів та процес розробки інформаційної системи, що представлена у вигляді комп’ютерної програми. Показано алгоритм визначення відповідності статистичного розподілу. Було розроблено програмну систему для статистичного аналізу розподілів глобальних брендів, основною особливістю якої є простота користування. Встановлено відповідності наборів даних глобальних брендів до закону розподілу Парето та до закону розподілу Ципфа.

# ABSTRACT

**Solohub V. A. Development of information system suited for statistical analysis of global brands distributions.** Specialty 122 «Computer science». Vasyl' Stus Donetsk National University, Vinnitsia, 2022.

This qualification work studies methods of statistical analysis of global brands distributions and development process of information system which is represented by computer program. Algorithm of estimation of correspondance to distribution law was shown. Correspondance of datasets (3) to Pareto Law and Zipf's Law were defined.

Key words: analysis, method, distribution, data, function, statistics, solution, program.

# ЗМІСТ

# ВСТУП

Більшість сучасних рішень для вирішення питань статистичного аналізу розподілів глобальних брендів потребує від користувача певних знань статистичного аналізу, математики та програмування. Дана робота націлена на вирішення питань часу та складності в даному контексті. Програмне забезпечення, розроблене у ході роботи, дозволить відкинути необхідність знання математики та програмування, залишив лише сферу статистичного аналізу. Це дозволить багатьом молодим брендам отримувати необхідну інформацію швидше та легше.

Дана робота присвячена вивченню методів статистичного аналізу та технологій парсингу даних **з метою** автоматизації процесу отримання розподілів брендів та з метою спрощення, пришвидшення й автоматизації статистичного аналізу цих розподілів.

**Завданням дослідження є**:

1. Визначити переваги та недоліки сучасних підходів до статистичного аналізу;
2. дослідити методи вивантаження даних з веб-сторінки;
3. дослідити методи статистичного аналізу для апроксимації параметрів розподілу, для аналізу на відповідність до розподілу;
4. розробити інформаційну систему, що спрощує та підвищує швидкість процесу статистичного аналізу розподілів глобальних брендів;
5. опрацювати кілька експериментальних розподілів за допомогою розробленої інформаційної системи.

**Об'єктом дослідження** є інформаційні системи для проведення статистичного аналізу розподілів брендів.

**Предметом дослідження** є веб-сторінка RankingTheBrands, методи парсингу AngleSharp та методи статистичного аналізу MathNet.

Дана розробка зможе надати можливість статистичного аналізу широкого списку різноманітних брендів у декілька натискань клавіші миші.

# 1 АНАЛІЗ ОБ’ЄКТУ ДОСЛІДЖЕННЯ ТА ДЕТАЛІЗАЦІЯ ЗАВДАНЬ

## 1.1 Базові терміни та визначення

За стандартом ISO 10668 **бренд** - це торгівельна марка, логотип або малюнок що описує загальні уявлення про продукт і має створювати чітке представлення товару в думках зацікавленої сторони та, таким чином, створювати економічні вигоди для власника.[15]

За українським законодавством **інформаційна система** це - організаційно-технічна система обробки інформації за допомогою технічних і програмних засобів.[1] Класичне визначення ІС визначає, що це система яка отримує інформацію на вхід, обробляє її, та видає на вихід відповідний результат.

За енциклопедією сучасної України **статистичний аналіз** – аналіз статистичних даних про масові явища та процеси в соціально-економічній сфері, природі, науці, техніці з метою встановлення закономірностей стану та розвитку цих явищ та процесів, зв'язку між ними, структурних зрушень, їх прогнозування.[2]

З практичної точки зору під терміном “**розподіл**” розуміють функцію, що описує відношення між спостереженнями в замкненому просторі. Будь який набір даних має свій розподіл.[3]

**Розподіл даних** це дані, що слідують певному закону розподілу.

## 1.2 Конкурентний аналіз систем статистичного аналізу розподілу даних

Зазвичай інформаційні системи для аналізу розподілу даних існують у вигляді пакетів, бібліотек для різноманітного програмного забезпечення.

Одним з таких рішень є **MathWorks MATLAB**. Дана програма надає можливість працювати з обширною кількістю законів розподілу. Робота з законами розподілу відбувається безпосередньо у консолі програми, де користувач вводить команди. Підбір параметрів нормального закону під розподіл даних у MATLAB виглядає таким чином:

```
x = grades(:,1);
```

Fit a normal distribution to the sample data by using `fitdist` to create a probability distribution object.

```
pd = fitdist(x,'Normal')
```

```
pd =
  NormalDistribution

  Normal distribution
       mu = 75.0083   [73.4321, 76.5846]
    sigma =  8.7202   [7.7391, 9.98843]
```

Рисунок 1.1 Підбір параметрів нормального закону розподілу за допомогою MATLAB

У результаті підстановки вважається що поданий набір даних x розподілений за нормальним законом розподілу, тоді параметрами цього розподілу є mu та sigma, що визначаються відповідно до закону розподілу. Усі методи викликаються мануально, від користувача очікується знання теорії ймовірностей. MATLAB також дає можливість роботи з функціями розподілу (CDF, PDF, EDF і так далі). Повний перелік знайти можна за джерелом.[4] Але дана програма не має функцій для перевірки даних на відповідность до законів розподілу. Користувачу доводиться робити це вручну.

Іншим рішенням у цій сфері є бібліотека **Python “scipy”**, що містить у собі методи побудови різноманітних розподілів даних. Для використання цієї бібліотеки спочатку слід встановити Anaconda Distribution. Anaconda Distribution - це масивна збірка бібліотек для роботи з даними у Python. SciPy.stats містить у собі методи для знаходження параметрів, підгонки, методи пошуку моментів, методи побудови **CDF**, **PDF**, також дозволяє обчислювати статистику **Колгоморова-Смирнова**, надає можливість робити **ресамплінг** (bootstrap). Але усе це відбувається у коді, тому використання цієї бібліотеки потребує знань Python. Також відсутні методи порівняння розподілів.[5]

Display the probability density function (pdf):

```
>>> x = np.linspace(norm.ppf(0.01),
...                 norm.ppf(0.99), 100)
>>> ax.plot(x, norm.pdf(x),
...        'r-', lw=5, alpha=0.6, label='norm pdf')
```

Рисунок 1.2 Підбір параметрів до Нормального закону розподілу з даних x та вивід на екран у Python scipy.stats

Останнє рішення, що було розглянуто - спеціалізована для статистичного аналізу мова програмування **R**. Це рішення в області статистичного аналізу так як і Python використовує пакети для роботи з розподілами даних. Кожен розподіл, що R підтримує, описується чотирма методами:[6]

1. p - для побудови **CDF**;
2. q - для побудови квантилю, тобто **протилежної CDF**;
3. d - для побудови **PDF**;
4. r - для побудови випадкового значення обраного розподілу.

Both of the R commands in the box below do exactly the same thing.

```
pnorm(27.4, mean=50, sd=20)
pnorm(27.4, 50, 20)
```

Рисунок 1.3 Отримання значення з Нормально розподіленого набору даних у точці x = 27.4, середнім = 50 і відхиленням 20

У R зазвичай для кожного закону розподілу є свій окремий пакет, наприклад для степеневих законів використовують **poweRlaw**. Даний пакет є ефективним рішення як для математично досвідчених користувачів оскільки надає велику свободу при обчислюванні, так і для звичайних користувачів, оскільки шлях до порівняння розподілів зовсім не довгий. Але все ж таки цей пакет потребує від користувача знань R.[7]

Можна сказати, що ПЗ конкурентів було розроблено не ставлячи за мету аналіз економічних розподілів. Користувач повинен сам усвідомлювати з якими даними та з якою метою він користується тою чи іншою програмою.

Таблиця 1.1 - Аналіз конкурентних розробок.

| | **Програмне Забезпечення** | | |
|---|---|---|---|
| **Критерій** | MATLAB | SciPy | R |
| Потреба технічних знань | наявна | наявна | наявна |
| Об’єм даних для потенційного аналізу | повний | повний | повний |
| Методи для порівняння розподілів даних | наявні | наявні | наявні й реалізовані (окремі пакети) |
| Взаємодія з даними | повна | повна | повна |

## 1.3 Деталізація завдань розробки

З критеріїв, запропонованих у таблиці 1.1, випливає, що сьогоденні рішення потребують від користувача значних знань в області статистичного аналізу та базових знань математики та програмування. Розглянуте ПЗ потребує від користувача увагу, вміння ефективно працювати з даними, розуміння принципів роботи з даними.

Тому у ході роботи необхідно розробити інформаційну систему, яка:

- Буде аналізувати розподіли вартостей глобальних брендів;
- не буде потребувати від користувача знань математики та програмування;
- буде простою у використанні;
- буде швидко приймати дані на вхід та швидко видавати результати в найпростішому вигляді.

З порівняння видно, що заплановане ПЗ повинно бути простим у користуванні та націленим на вирішення лише певної проблеми. Задля досягнення цього, було вирішено відкинути відкритість даних.

# 2 РОЗРОБКА ІНФОРМАЦІЙНОЇ СИСТЕМИ

## 2.1 Аналіз інформаційних ресурсів з необхідними даними

Для написання ПЗ було обрано мову програмування **Visual C#**. Вибір саме цієї мови полягає у тому, що це розробка Microsoft, тому вона буде підтримуватись достатньо довго та вона підтримує платформу .NET, що дає змогу їй запускатись на великій кількості різних комп’ютерів.

У ході дослідження було поставлено за мету розробити систему для аналізу розподілів вартостей брендів таким чином, щоб вона дозволяла швидко звантажувати та обробляти дані. Для цього було розглянуто результати тестування п’ятьох бібліотек для **парсингу веб-сторінок** у C#:

1. HtmlAgilityPack;
2. Fizzler;
3. CsQuery;
4. AngleSharp;
5. Regex.

Тестування швидкодії проводили на одному й тому самому наборі даних. При розрахуванні часу намагались відкинути все, що напряму не завантажує дані.

Набір даних - локально завантажена сторінка, з неї витягували всі дані з тегом “href”. [9]

Серед розглянутих варіантів було обрано **AngleSharp**, оскільки саме ця бібліотека надає можливість завантажити сторінку в форматі HTML та обрати необхідні теги для розбору даних.  Дана бібліотека має баланс між простотою використання та швидкодією.

Для роботи з даними було обрано бібліотеку **MathNet**. Ця бібліотека надає інструменти для статистичного аналізу розподілу даних.

Для роботи з графічним представлення було обрано **ScottPlot**. Ця бібліотека надає інструменти для представлення даних у вигляді графіків.

## 2.2 Розробка архітектури системи

У ході роботи було створено проект у Visual Studio на платформі .NET 6.0. Далі було додано форми:

- Список listBox, що містить набори даних;
- кнопка connectButton, що підвантажує дані з мережі;
- кнопка downloadPageDatasetButton, що завантажує дані обраного набору;
- кнопка analyzePLButton, що запускає процес статистичного аналізу та перевіряє дані на відповідність до обраного розподілу;
- текстове вікно outputTextBox, що виводить результати роботи функцій;
- перемикач numericBootstraps, що визначає кількість ітерацій при перевірці на відповідність до закону розподілу;
- список distributionTypeComboBox, що визначає розподіл, з яким буде порівнюватися обраний набір даних;
- текстова мітка datasetLabel, що показує завантажений на даний момент набір даних.

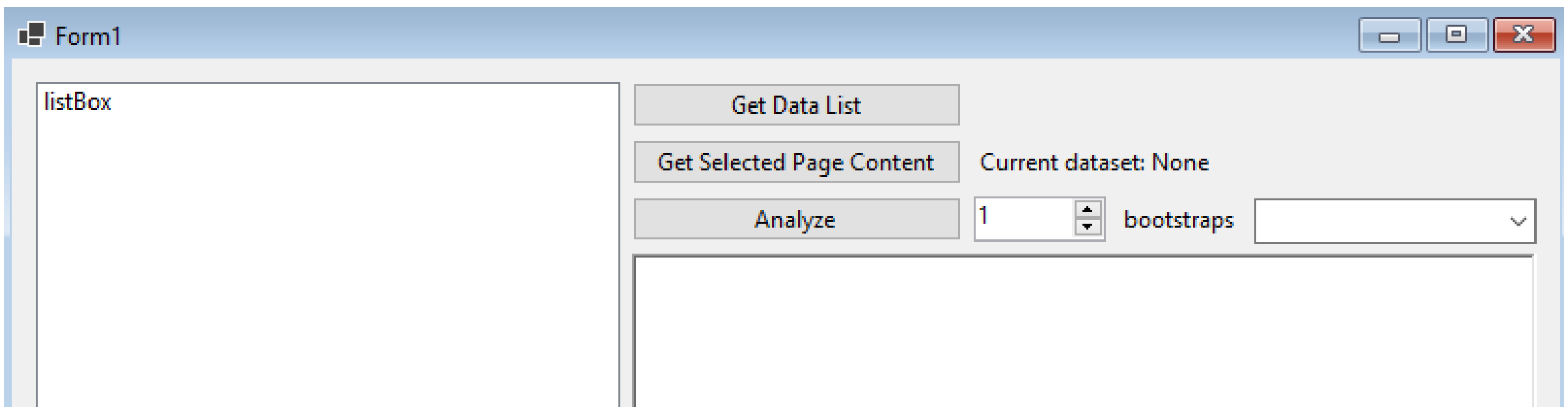

Рисунок 2.1 Вигляд користувацького інтерфейсу ПЗ

## 2.3 Обґрунтування інструментарію розробки

За дослідження необхідно було визначити які саме закони розподілу будуть використовуватись для порівняння з експериментальними даними, отриманими в ході роботи програми. Для цього необхідно розібрати підхід до статистичного аналізу розподілу глобальних брендів. Це було зроблено для списку ТОП 100 найкращих світових брендів Best Global Brands, що публікується Interbrand щороку.

За дослідження було використано R та пакети poweRlaw і ptsuit що дають інструменти для перевірки розподілу даних на відповідність до Степеневого закону розподілу та до закону Ципфа відповідно. Обидва порівняння було зроблено за допомогою методу Bootstrap при 500 ітераціях, що дозволяє ітераційно порівняти відстані (критерії Колмогорова-Смирнова) розподілу ймовірностей даних та моделі ймовірностей цього розподілу даних підігнану під відповідний закон розподілу з метою отримання ймовірності відповідності - p-value. Для кожної ітерації, при точності = 0,1, перевірялась така гіпотеза:

- $H_0$ : змінна bv20** належить до розподілу моделі;
- $H_1$ : змінна bv20** НЕ належить до розподілу моделі.

У ході роботи було отримано таку таблицю ймовірностей:

Таблиця 2.1 - ймовірності відповідності до законів розподілу.

| Змінна моделі | p-value Power Law | p-value Zipf |
|---|---|---|
| bv2010 | 0,65 | 3,512e-08 |
| bv2011 | 0,01 | 4,047e-07 |
| bv2012 | 0,16 | 5,522e-06 |
| bv2013 | 0,2 | 5,552e-06 |
| bv2014 | 0,33 | 5,564e-08 |
| bv2015 | 0,05 | 5,042e-05 |
| bv2016 | 0,36 | 8,036e-07 |
| bv2017 | 0,12 | 5,855e-07 |
| bv2018 | 0,32 | 9,123e-06 |
| bv2019 | 0,39 | 5,865e-11 |
| bv2020 | 0,21 | 2,267e-06 |

Результати перевірки показали, що більшість розподілів даних ймовірно розподілено за степеневим законом розподілу. Відповідно обидві перевірки буде реалізовано у програмному забезпечені, оскільки обидва розподіли тісно пов’язані.

## 2.4 Модуль видобутку даних

Будь-який закон розподілу має кілька виглядів. Одним з цих виглядів є CDF функція розподілу. **Функція розподілу ймовірностей (CDF)** повідомляє про відсоток даних n менше всіх можливих n.[10] Наявність розподілу даних в такому вигляді надає можливість порівняти цей розподіл з його моделлю. Таке порівняння називають **статистикою Колмогорова-Смирнова**. Воно відбувається за такою формулою[7]:

$$D(x) = \max_{x \geq x_{min}} |F(x) - F_n(x)|$$

де $F(x)$ – значення CDF розподілу даних в точці $x$;

$F_n(x)$ – значення CDF моделі розподілу в точці $x$;

$D(x)$ – статистика Колмогорова-Смирнова для значення набору даних $x$.

При перевірці статистика Колмогорова-Смирнова використовується для пошуку параметру ***shape*** обидвох необхідних законів розподілу та для пошуку ймовірності відповідності ***p-value***.

Закон Ципфа - емпіричний закон розподілу, при якому, частота появи елемента пропорційна його порядковому номеру[14]. Наприклад маємо елемент *max*, що зустрічається найчастіше у списку, тоді другий, найчастіше зустрічний, елемент буде зустрічатись у два рази менше ніж *max*, третій - у три рази менші ніж *max*, і т.д. Має один параметр ***shape(xmin)***.

Степеневий закон базується на принципі Парето, та стверджує, що 80% явищ спричинені 20% причин[11]. Для наших розподілів даних вірна така постановка: “на 20% найдорожчих брендів рейтингу зводиться 80% коштів”. Справедлива і протилежна постановка: “на 80% найдешевших брендів рейтингу зводиться 20% коштів”. Степеневий закон формує степеневий розподіл. Задається двома параметрами ***alpha(scale)*** та ***xmin(shape)***.

Для порівняння розподілу даних на відповідність до степеневих законів користуються таким алгоритмом дій[7] (АЛГОРИТМ 1):

1. Нехай експериментальний розподіл *pareto* створено за Степеневим законом, тоді початковий параметр *xmin = min(pareto)*;
2. Знаючи параметр *xmin*, можемо знайти параметр *alpha*[12]:

$$\alpha = 1 + n \cdot \left[\sum_{i=1}^{n} \ln\left(\frac{x_i}{x_{min}}\right)\right]^{-1}$$

3. Знайшовши *alpha* можна знайти наближення до *xmin*:

3.1. Знаходимо статистику Колмогорова-Смирнова *D* для *n* значень *pareto*;

3.2. Наше *xminEst* є значенням *x*, що мінімізує *D*;

4. Маючи параметри, можемо проаналізувати відповідність розподілу даних до закону розподілу. Обчислюємо статистику КС для оригінального набору даних, отримуємо змінну *KSTd*;

5. Створюємо *n1*, що дорівнює кількості значень менше *xminEst*;
6. Створюємо *n2 = n – n1*, та *P = 0*;
7. Тоді запускаємо Бутстреп на *B* ітерацій:

 7.1. Симулюємо пов'язану *KSTsim* статистику КС для моделі *n1* значень Uniform розподілу з параметрами 1, *xminEst* і для моделі *n2* значень з Power law розподілу;

 7.2. Якщо *KSTd* > *KSTsim*, тоді *P++*;

8. Після завершення знаходимо *p-value = P/B*;

Даний підхід доречно використовувати для будь-якого закону розподілу, при розрахунках лише необхідно міняти відповідні моделі та їх функції розподілу ймовірностей.[13]

Для завантаження та парсингу даних зі сторінки було створено клас DataParse. Він складається з восьми полів та шести методів.

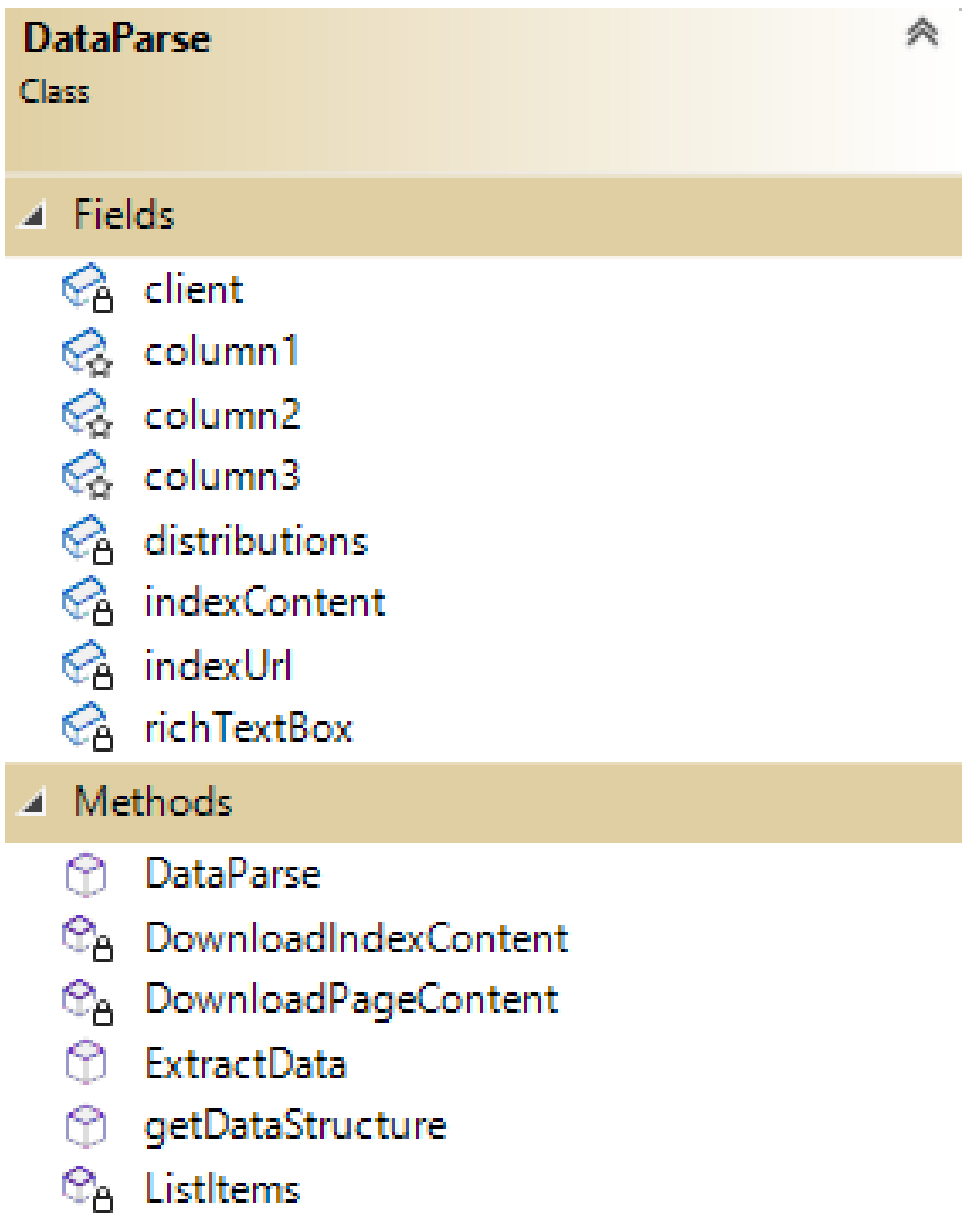

Рисунок 2.2 Діаграма класу DataParse

Конструктор DataParse(ListBox, RichTextBox) на вході отримує посилання на дві форми:

- ListBox, що відповідає за відображення та обирання наборів даних для аналізу;
- RichTextBox, що відображає результати дій користувача.

Задача створеного конструктора - завантажити головну сторінку indexUrl, та створити список наборів даних з їх назвою.

```
public DataParse(ListBox ListBox, RichTextBox RichTextBox)
{
    client = new HttpClient();
    distributions = new Dictionary<string, string>();
    richTextBox = RichTextBox;
    DownloadIndexContent(ListBox);
}
```

Рисунок 2.3 Реалізація конструктору DataParse()

За роботи конструктору для цього викликається метод async DownloadIndexContent(). Цей метод є технічною реалізацією рішення поставленої задачі. Це асинхронний метод, що завантажує, відповідне за текстовий вміст сторінки, поле indexContent за допомогою методу бібліотеки System.Net.Http.HttpClient.ReadStringAsync().

```
private async void DownloadIndexContent(ListBox listBox) //get index
{
    HttpResponseMessage response = await client.GetAsync(indexUrl);
    response.EnsureSuccessStatusCode();
    indexContent = await response.Content.ReadAsStringAsync();
    ListItems(listBox);
    Debug.WriteLine("Downloaded Datalist");
    richTextBox.Text += "Downloaded Datalist successfully";
}
```

Рисунок 2.4 Реалізація методу DownloadIndexContent()

У створеному методі викликається функція async ListItems(), що парсить дані з сайту у, вказаний у конструкторі, ListBox. Із завантаженої сторінки indexContent обираються усі поля з тегом a за класом “listRankings” та з тегом span за класом “rankingName”, далі запускається цикл, що заповнює словник distributions та ListBox. Після виконання цього методу, користувач отримує повідомлення про успішне завантаження списку даних.

```
private async void ListItems(ListBox listBox) //get distribution list
{
    var config = Configuration.Default;
    using var context = BrowsingContext.New(config);
    using var doc = await context.OpenAsync(req => req.Content(indexContent));

    var aAll = doc.QuerySelectorAll("a.listRankings");//get all "a" with class = "listRankings"
    var spans = doc.QuerySelectorAll("span.rankingName");

    for (int i = 0; i < spans.Count(); i++)
    {
        distributions.Add(spans.ElementAt(i).TextContent, aAll.ElementAt(i).GetAttribute("href"));
        listBox.Items.Add(spans.ElementAt(i).TextContent);
    }
}
```

Рисунок 2.5 Реалізація методу ListItems()

У формі цей метод викликається після натискання кнопки connectButton. При цьому користувачу надається доступ до кнопки відповідальної за завантаження наборів даних. Якщо сторінка будь-яким чином не завантажилась, користувач отримує помилку та повідомлення про помилку.

```
private void connectButton_Click(object sender, EventArgs e)
{
    try
    {
        parser = new DataParse(listBox, outputTextBox);
        connectButton.Enabled = false;
        downloadPageDatasetButton.Enabled = true;
    }
    catch (Exception ex)
    {
        outputTextBox.Text = ex.Message;
        outputTextBox.Text += "\nCould not connect to RankingTheBrands.com";
    }
}
```

Рисунок 2.6 Реалізація події connectButton_Click()

Після виклику конструктора користувачу доведеться взаємодіяти з класом ExtractData(). Цей клас відповідає за завантаження обраного, у ListBox, набору даних та за підгонку цих даних в універсальний формат. Усі завантажені дані зберігаються у списку словників, де ключ - це назва бренду, а значення - кортеж з року дослідження, рангу бренду в датасеті та числового значення. Початково дані завантажуються у форматі тексту, оскільки виконання цього методу для великих наборів є часомістким процесом. Цей метод виводить набір даних та зберігає їх формат у полях column1, column2, column3. У ході роботи методу виділяється вміст першої комірки цільової сторінки з тегом div за класом top100row. З цього вмісту обираються теги відповідні за ранг, назву бренду, значення.

```
var div = doc.QuerySelector("div.top100row"); //get first row

var rank = div.QuerySelector("div.pos"); //get first column value
var name = div.QuerySelector("div.name"); //get second column value
var value = div.QuerySelector("div.weighted"); //get third column value
```

Рисунок 2.7 Теги, що визначають формат даних

Отримавши значення вирішується формат. Жирним наведено кінцевий формат, звичайний текст відповідає за змінні та їх параметри. Необхідність перевіряти заповненість комірки рангу пояснюється тим, що на сайті RankingTheBrands є набори даних з наявним місцем під ранг, та відсутнім там значенням рангу.

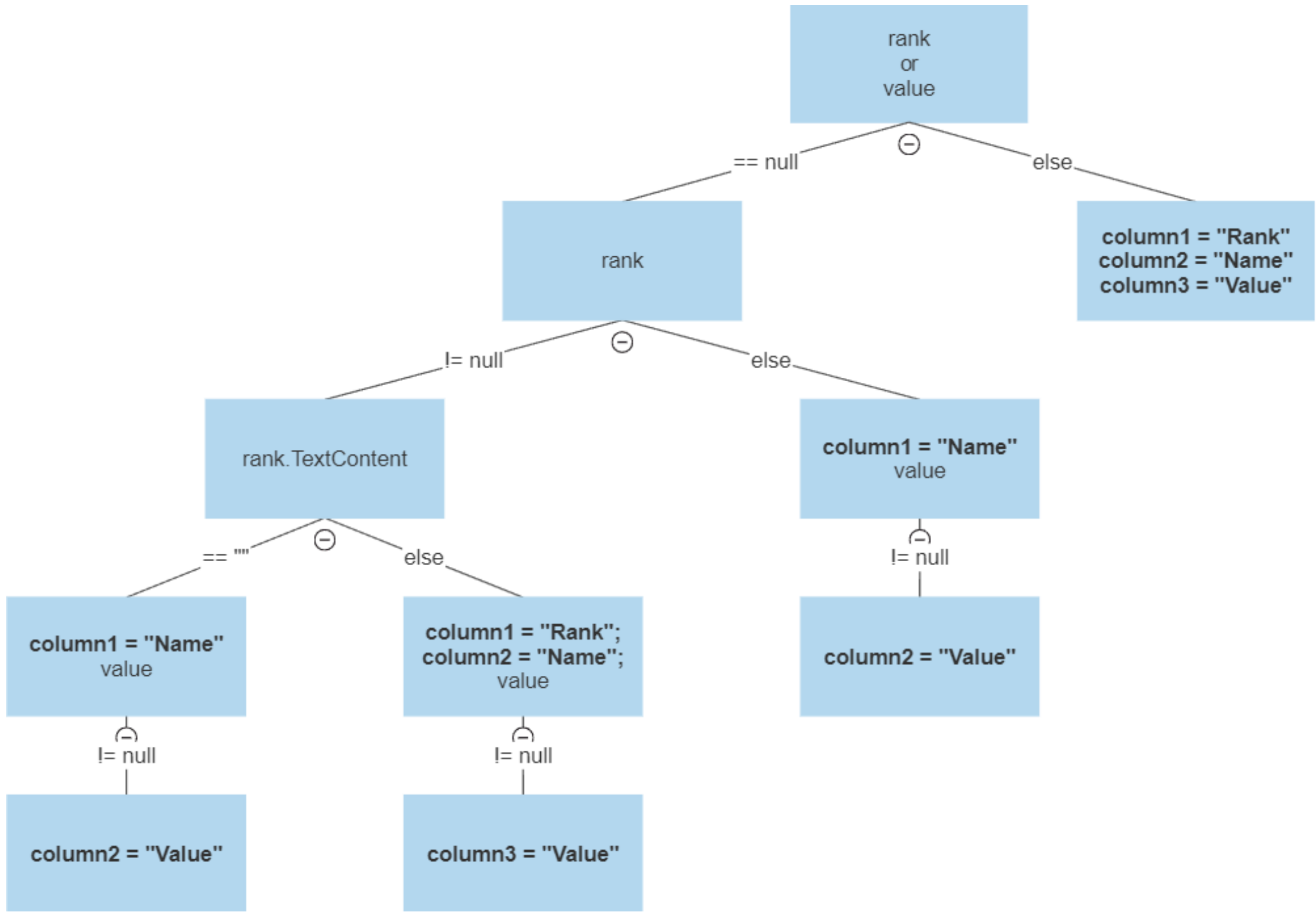

Рисунок 2.8 Логіка вирішення формату даних (створено за допомогою smartdraw.com)

Після необхідно виділити список доступних років. Обирається тег select, що містить поле параметр name. З отриманого списку обираються усі значення с тегом option.

```
var select = doc.QuerySelector("select[name]");//get year list
var options = select.QuerySelectorAll("option");//get year id
```

Рисунок 2.9 Теги, що визначають кількість років (окремих списків)

Відповідні сторінки завантажуються

```
foreach (var option in options)
{
    yearList.Add(option.TextContent);
    datasetPages.Add(await DownloadPageContent(distributions[listBox.GetItemText(listBox.SelectedItem)]
        + "&year=" + option.GetAttribute("value")));//Download all years pages
    Debug.WriteLine(option.TextContent);
    Debug.WriteLine(option.GetAttribute("value"));
    richTextBox.Text += "\n" + option.TextContent;
}
```

Рисунок 2.10 Код, що відповідає за завантаження текстової частини сторінок за кожен рік

Далі списки підганяються у відповідний формат даних. Для кожної завантаженої сторінки обираються усі рядки з даними, далі кортеж заповнюється згідно формату. У кінці усі завантажені дані йдуть до словника, отриманий словник додаємо до списку. Це дозволяє працювати з даними за кожен рік. Після цього метод виводить список і завершує свою роботу.

```
if (column1 == "Name")...
else if (column1 == "Rank" && column2 == "Name" && column3 != "Value")...
else if (column1 == "Name" && column2 == "Value")...
else if (column1 == "Rank" && column2 == "Name" && column3 == "Value")...
else//if data representation on website is broken in some way
{
    Debug.WriteLine("Failed to form a table");
    richTextBox.Text += "\nFailed to form a table";
}
return dataset;
```

Рисунок 2.11 Логіка, відповідальна за підгон

## 2.5 Модуль обробки даних

Для роботи з даними було створено клас DataAnalysis. Він складається з 9 методів.

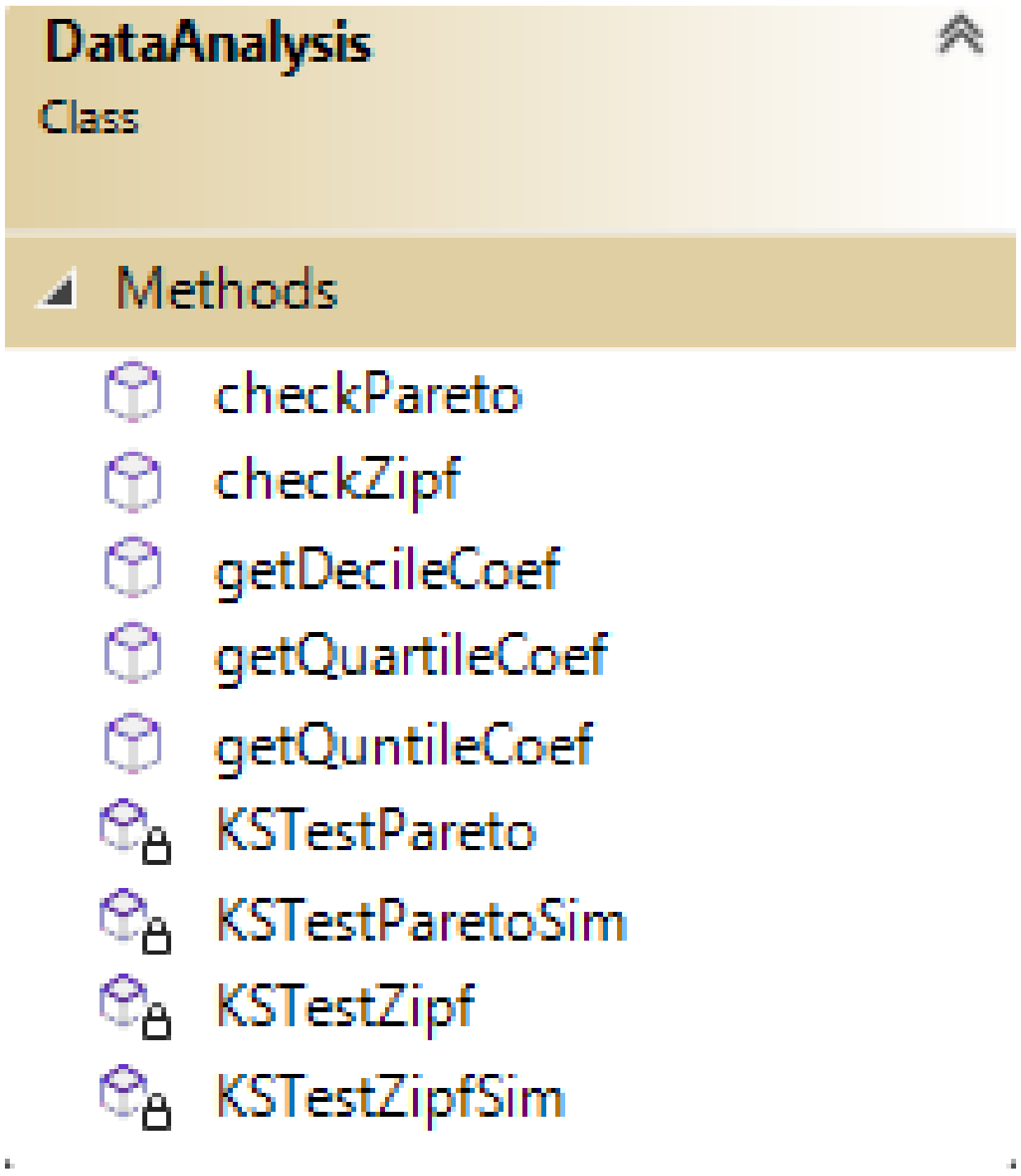

Рисунок 2.12 Діаграма класу DataAnalysis

Основна взаємодія користувача з цим класом у формі відбувається за допомогою кнопок analyzeButton, downloadPageDatasetButton, списку listBox, списку distributionTypeComboBox, лічильника numericBootstraps та прапорцю coefCheckBox.

Методи checkPareto та checkZipf побудовані за алгоритмом 1. До checkPareto виставляється така гіпотеза:

- $H_0$ - розподіл даних подібний до розподілу Парето;
- $H_1$ - розподіл даних НЕ подібний до розподілу Парето.

До checkZipf[15] виставляється така гіпотеза:

- $H_0$ - розподіл даних побудовано з розподілу Ципфа;
- $H_1$ - розподіл даних НЕ побудовано з розподілу Ципфа.

На вхід методи отримують набір даних, його структуру, кількість ітерацій та посилання на вікно виводу. Робота цих методів починається з конвертації даних з текстових до числових. За цього процесу також обирається параметр xmin це найменше значення у розподілі. У випадку закону Ципфа, xmin це константа та дорівнює одиниці.

```
foreach (var year in dataset)//foreach year in dataset
{
    Dictionary<int, double> ranksValues = new Dictionary<int, double>();
    foreach (var value in year.Values)//foreach value in current year
    {
        int item1;
        double item2;
        int.TryParse(value.Item1, NumberStyles.Any, CultureInfo.InvariantCulture, out item1);//t
        Double.TryParse(value.Item2, NumberStyles.Any, CultureInfo.InvariantCulture, out item2);
        //Debug.WriteLine("rank: " + item1 + ", which has type" + item1.GetType());
        //Debug.WriteLine("value: " + item2 + ", which has type" + item2.GetType());
        ranks.Add(item1);
        ranksValues.Add(item1, item2);
    }
    try
    {
        xmin.Add(ranksValues.Values.Min());
    }
```

Рисунок 2.13 Перетворення тексту на числа, пошук xmin на прикладі методу checkPareto()

Далі, за методом MLE для степеневих законів розподілу обчислюється параметр alpha, причому він не є наближенням.

```
List<double> alpha = new List<double>(); //getting alpha by
int i = 0;
foreach (var year in ranksValuesList)
{
    double sum = 0;
    foreach (var xi in year.Values)
    {
        sum += Math.Log(xi / xmin[i]);
    }
    i++;
    alpha.Add(1 + year.Values.Count() * Math.Pow(sum, -1));
}
```

Рисунок 2.14 Пошук параметру alpha

Після параметри виводяться у вікно виводу;

```
for (int j = 0; j < xmin.Count(); j++)
{
    Debug.WriteLine("xmin" + j + ": " + xmin[j]);
    rtb.Text += "\n" + "xmin" + j + ": " + xmin[j];
    Debug.WriteLine("alpha" + j + ": " + alpha[j]);
    rtb.Text += "\n" + "alpha" + j + ": " + alpha[j];
}
```

Рисунок 2.15 Вивід

Потім, за допомогою статистики Колмогорова-Смирнова, знаходиться наближення xminEst. Індекс числа *x*, при якому значення функції *D(x)* мінімальне - є індексом нашого xminEst.

Знайшовши xminEst, визначаються $n_1$, $n_2$. Далі знаходяться статистики Колмогорова-Смирнова для обраного набору даних при параметрах розподілу alpha, xminEst (у випадку закону Ципфа, параметри alpha та xmin = 1). Значення статистик передаються до вікна виводу.

Запускається процес Бутстрепінгу. За попередньо вказаною кількістю ітерацій симулюється значення статистик моделі обраного розподілу із визначеними параметрами xminEst, alpha для розподілу Парето та параметрами alpha, xmin = 1 для розподілу Ципфа. При кожній ітерації порівнюється чи є значення КС статистики для набору даних більшим за отримане в ході симуляції значення КС статистики моделі, у результаті успіху такого порівняння значення змінної P збільшується на одиницю. Усі моделі симулюються відповідно параметру xminEst та випливаючих з нього кількостей $n_1$ та $n_2$.

```
for (int I = 0; I < B; I++)//starting bootstrapping
{
    List<double> KSTsim = KSTestParetoSim(ranksValuesList, xminEst, alpha, n1, n2);
    /*foreach (var D in KSTsim)
    {
        Debug.WriteLine("KSTsim = " + D);
    }*/
    for (int N = 0; N < KSTd.Count(); N++)//foreach statistic
    {
        if (KSTd[N] > KSTsim[N])
        {
            PList[N]++;//increment each P when KSTd > KSTsim
        }
    }
}
```

Рисунок 2.16 Симуляція значень KSTsim

Після завершення процесу Бутстрепінгу в B ітерацій для кожного року набору даних, обраховується p-value, що є часткою P на B для кожного року. Усі дії відбуваються згідно дослідженому алгоритму.

```
foreach(var P in PList)
{
    p_values.Add(P/B);
}
```

Рисунок 2.17 Отримання результату

За вказівок керівника наукової роботи також необхідно було реалізувати обчислення децильних, квантильних та квартильних коефіцієнтів. Відповідні методи називаються getDecileCoef(), getQuantileCoef(), getQuartileCoef(). Вони відповідають за знаходження відношення певної частки найдорожчих брендів до певної частки найдешевших. Робота цих методів починається з пошуку n, кількості років. Далі визначається яка частка для поданого набору даних буде квантилем, квартилем або децилем

```
List<double> decileCoefficients = new List<double>();
int count = ranksValuesList[0].Values.Count();
int yearCount = ranksValuesList.Count();
int decile = (int)Math.Round(count * 0.1);
```

Рисунок 2.18 Пошук дециля для будь якого набору даних

Отримавши значення дециля знаходяться суми першого та останнього. За дослідження було визначено, що дані на веб-сторінці представлено у сортованому вигляді, тому можна одразу рахувати суми.

```
for (int i = 0; i < yearCount; i++)
{
    topSum.Add(0);
    for (int j = 0; j < decile; j++) //this works because ea
    {
        topSum[i] += ranksValuesList[i].ElementAt(j).Value;
    }
}

for (int i = 0; i < yearCount; i++)
{
    botSum.Add(0);
    for (int j = count - decile; j < count; j++) //this work
    {
        botSum[i] += ranksValuesList[i].ElementAt(j).Value;
    }
}
```

Рисунок 2.19 Знаходження верхньої та нижньої сум

Після цього обчислюється частка цих сум, результати виводяться на екран.

```
for (int i = 0; i < yearCount; i++)
{
    decileCoefficients.Add(topSum[i] / botSum[i]);
    rtb.Text += "\nDecile coefficient " + i + " = " + decileCoefficients[i];
}
```

Рисунок 2.20 Пошук децильних коефіцієнтів

## 2.6 Модуль взаємодії з даними

Усі ці методи передаються у форму та в її івенти. Виділеним позначаються поля та методи описані у коді.

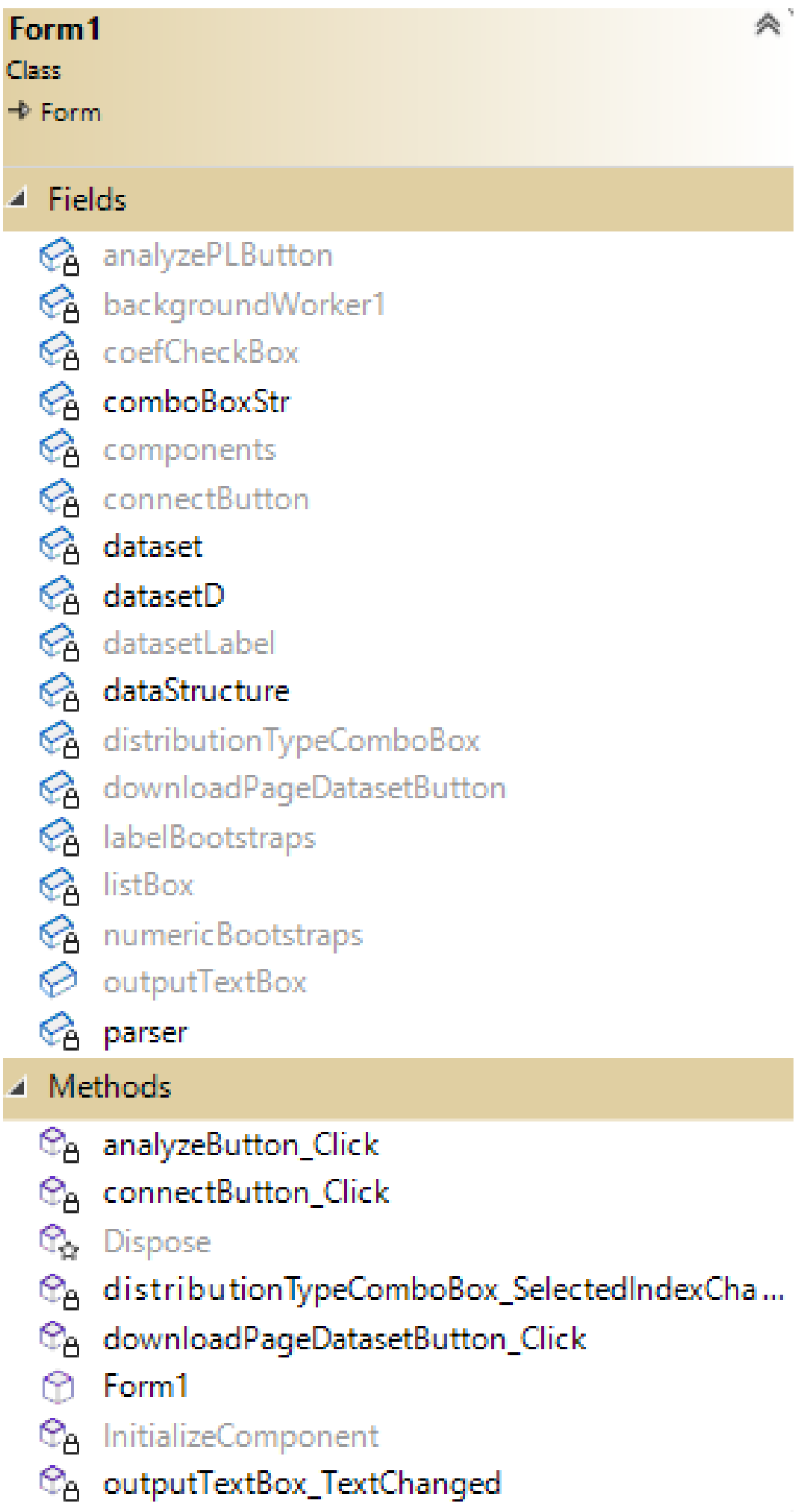

Рисунок 2.21 Діаграма класу Form1

При ініціалізації для форми створюються поля відповідальні за перемикач розподілів, збереження та завантаження даних. Для роботи з класом DataParse створюється об'єкт класу. Його перший виклик відбувається за івенту connectButton_Click. Ця подія відповідає за спробу підключення до сторінки. Також йде ініціалізація об'єкту parser.

```
{
    try
    {
        parser = new DataParse(listBox, outputTextBox);
        connectButton.Enabled = false;
        downloadPageDatasetButton.Enabled = true;
    }
    catch (Exception ex)
    {
        outputTextBox.Text = ex.Message;
        outputTextBox.Text += "\nCould not connect to RankingTheBrands.com";
    }
}
```

Рисунок 2.22 Перевірка підключення до сайту RankingTheBrands

У випадку помилки, програма повідомить, що вона не змогла завантажити дані зі сторінки.

Після успішного підключення, користувач побачить список датасетів. Виділивши один з них та натиснувши кнопку завантаження, він розпочне подію downloadPageDatasetButton_Click. Її задача - отримати формат даних та сформувати відповідний набір даних. У вікні вивдоу видаляються усі минулі повідомлення та викликається асинхронний метод, ExtractData. Кнопки блокуються. Програма пробує виконати задачу по виконанню дії datasetD. Обраний набір даних завантажується у пам'ять. Дочекавшись виконання, кнопки вмикаються. Інакше, програма виводить повідомлення про помилку.

```
private async void downloadPageDatasetButton_Click(object sender, EventArgs e)
{
    if (listBox.SelectedItem != null)
    {
        downloadPageDatasetButton.Enabled = false;
        analyzePLButton.Enabled = false;
        outputTextBox.Text = "";
        datasetD = parser.ExtractData(listBox, datasetLabel);
        try
        {
            dataset = await datasetD;
        }
        catch (Exception ex)
        {
            outputTextBox.Text = ex.Message;
        }
        downloadPageDatasetButton.Enabled = true;
        analyzePLButton.Enabled = true;
    }
    else
    {
        Debug.WriteLine("Select item");
        outputTextBox.Text += "\nSelect item";
    }
}
```

Рисунок 2.23 Завантаження даних асинхронним методом

При успішному виконанні у перший раз, програма вмикає кнопку аналізу analyzeButton. Програма також повідомляє про обраний розподіл. Далі користувачу надається вибір:

- Можливість обрати закон розподілу для аналізу на нього обраних даних (відбувається при виконанні події distributionTypeComboBox_SelectedIndexChanged);
- можливість, при аналізі, вивести динаміки коефіцієнтів кореляції натиснувши прапорець;
- можливість задати кількість ітерацій для симуляції.

Визначившись з налаштуваннями користувач повинен натиснути кнопку аналізу. При цьому викликається подія analyzeButton_Click. Ця подія перевіряє структуру даних на повноту, після цього обирає згідно запиту методи до виконання за допомогою структури switch-case.

```
case "Pareto law":
    {
        List<Dictionary<int, double>> data = DataAnalysis.checkPareto(dataset, dataStructure,
            Convert.ToInt32(numericBootstraps.Value), outputTextBox);//Convert.ToInt32(Math.Ro
        if(coefCheckBox.Checked)
        {
            DataAnalysis.getDecileCoef(data, outputTextBox);
            DataAnalysis.getQuntileCoef(data, outputTextBox);
            DataAnalysis.getQuartileCoef(data, outputTextBox);
        }
        break;
```

Рисунок 2.24 Випадок Pareto Law

# 3 ЕКСПЕРИМЕНТАЛЬНІ ДОСЛІДЖЕННЯ

## 3.1 Дослідження розподілу вартостей брендів Best German Brands за 2014-2015 роки

Дослідимо розподіли даних списку Best German Brands що містить спостереження найдорожчих німецьких брендів за 2014 та 2015. Кожен рік містить по 50 спостережень. Аналіз років йде у спадному порядку (параметри 0 відповідні спостереженням 2015 року, параметри 1 - спостереженням 2014 року).

При тестуванні на відповідність до закону Парето користувач отримує таке повідомлення:

```
Analyzing for Pareto distribution
xmin0: 150
alpha0: 1,446210114328602
xmin1: 102
alpha1: 1,3862785560650406
MIN D VALUE 0 = 0,001021850893238918; D VALUE INDEX = 6
Estimated XMIN0  = 7011
MIN D VALUE 1 = 0,0009890251891371271; D VALUE INDEX = 13
Estimated XMIN1  = 3036
n1 0 = 42
n1 1 = 36
n2 0 = 8
n2 1 = 14
Bootstrapping for 1000 iterations
KSTd 0 = 259,1230275017752
KSTd 0 = 109,75149915149959
p-value: 0,428
p-value: 0,311
Decile coefficient 0 = 97,26179775280899
Decile coefficient 1 = 124,46366279069767
Quantile coefficient 0 = 49,77919407894737
Quantile coefficient 1 = 54,51914098972922
Quartile coefficient 0 = 40,244505494505496
Quartile coefficient 1 = 42,388482023968045
```

Рисунок 3.1 Статистичний аналіз даних на відповідність до Pareto Law для розподілу Best German Brands

За умови, що точність = 0,1, можна сказати, що протестований розподіл даних є подібним до розподілу даних за законом Парето.

Протестуємо даний набір даних на закон Ципфа:

```
Analyzing for Zipf distribution
xmin0: 1
alpha0: 1,1378980918442434
beta0: 0,1378980918442434
xmin1: 1
alpha1: 1,1386236136854655
beta1: 0,1386236136854655
n1 0 = 1
n1 1 = 1
n2 0 = 49
n2 1 = 49
Bootstrapping for 1000 iterations
WARNING: This might take a long time
KSTd 0 = 0,659022371197671
KSTd 0 = 0,6317280655117823
p-value: 0
p-value: 0
Decile coefficient 0 = 97,26179775280899
Decile coefficient 1 = 124,46366279069767
Quantile coefficient 0 = 49,77919407894737
Quantile coefficient 1 = 54,51914098972922
Quartile coefficient 0 = 40,244505494505496
Quartile coefficient 1 = 42,388482023968045
```

Рисунок 3.2 Статистичний аналіз даних на відповідність до Zipf's Law для розподілу Best German Brands

За умови, що точність = 0,1, можна сказати, що протестований розподіл даних побудовано з інформації, яка не є проявом закону Ципфу.

## 3.2 Дослідження розподілу вартостей брендів Best Pharma Brands за 2016 рік

Дослідимо розподіли даних списку Best Pharma Brands що містить спостереження найдорожчих фармацевтичних брендів за 2016 рік. У даному наборі 10 спостережень.

При тестуванні на відповідність до закону Парето користувач отримає таке повідомлення:

```
Analyzing for Pareto distribution
xmin0: 6,778
alpha0: 2,6712845208064966
MIN D VALUE 0 = 0,014043868059830822; D VALUE INDEX = 8
Estimated XMIN0  = 8,123
n1 0 = 1
n2 0 = 9
Bootstrapping for 1000 iterations
KSTd 0 = 0,9806036335417935
p-value: 0,026
```

Рисунок 3.3 Тест на закон Парето для Best Pharma Brands

За умови, що точність = 0,1, можна сказати, що протестований розподіл даних не є подібним до розподілу даних за законом Парето.

Протестуємо даний набір даних на закон Ципфа:

```
Analyzing for Zipf distribution
xmin0: 1
alpha0: 1,3980853307242602
beta0: 0,3980853307242602
n1 0 = 1
n2 0 = 9
Bootstrapping for 1000 iterations
WARNING: This might take a long time
KSTd 0 = 0,8833560974871298
p-value: 0,315
```

Рисунок 3.4 Тест на закон Ципфа для Best Pharma Brands

За умови, що точність = 0,1, можна сказати, що протестований розподіл даних побудовано з інформації, яка є проявом закону Ципфу. Але не можна сказати, що цей набір даних можна віднести до закону Ципфа, оскільки це не відповідає гіпотезі й є неможливим взагалі.

### 3.3 Дослідження розподілу вартості брендів Best Global Brands за 2007-2021 роки

Дослідимо розподіли даних списку Best Global Brands що містить спостереження найдорожчих брендів світу за період 2007-2021 роки. У даному наборі по 100 спостережень за рік. Аналіз років йде у спадному порядку (параметри 0 відповідні спостереженням 2021 року, параметри 1 - спостереженням 2020 року і так далі...).

При тестуванні на відповідність до закону Парето жоден з наборів даних не задовільнив гіпотезі при точності 0,1, тому можна сказати, що протестований розподіл даних не є подібним до розподілу даних за законом Парето.

При тестуванні на відповідність до закону Ципфа жоден з наборів даних не задовільнив гіпотезі при точності 0,1, тому можна сказати, що протестований розподіл даних побудовано з інформації, яка не є проявом закону Ципфу.

Щоб пояснити невідповідність до протестованих законів, необхідно розглянути таблицю 2.1 ще раз. У таблиці наведено результати при тестуванні на відповідність до закону Ципфа та на відповідність до степеневого закону. Даний датасет відноситься до якогось степеневого закону розподілу, але це точно не закон Парето. Перевірка на відповідність до закону Ципфа також показала, що даний набір даних не схожий до розподілу даних за Ципфом. Тому цей набір даних, скоріше за все, відноситься до іншого степеневого розподілу.

Також отримали результати розрахунків децильних, квантильних і квартильних коефіцієнтів:

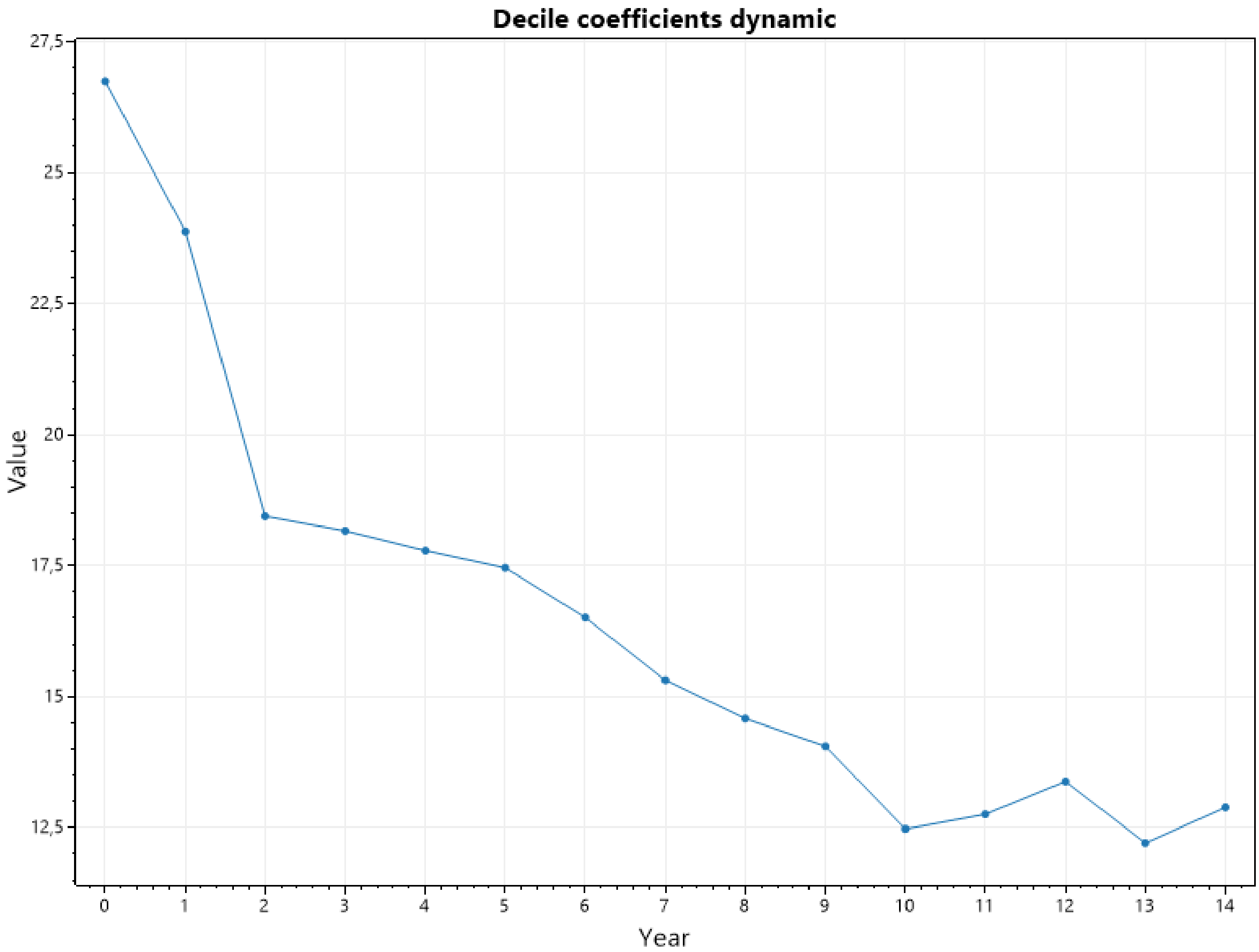

Рисунок 3.5 Децильні коефіцієнти даних Best Global Brands

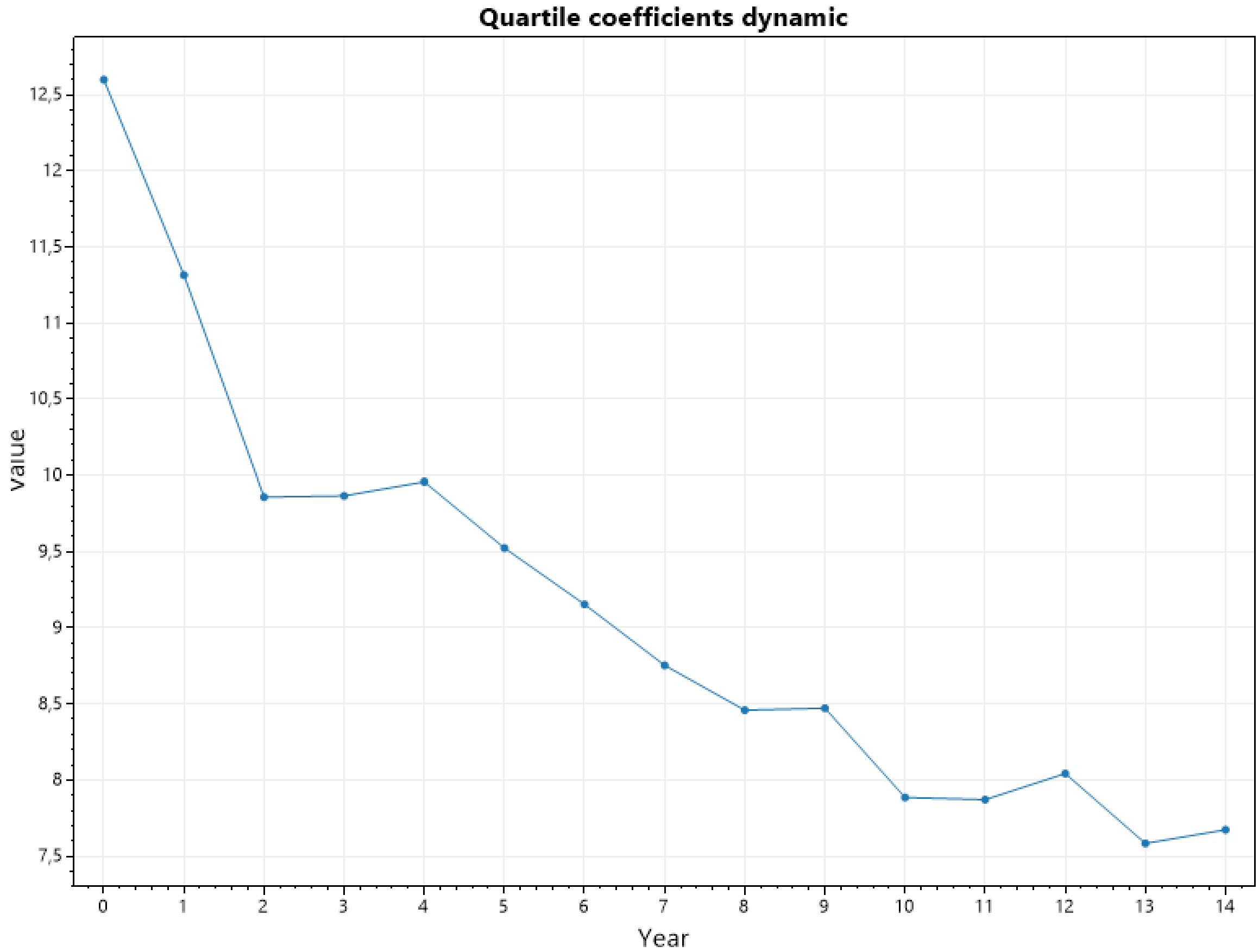

Рисунок 3.6 Квартильні коефіцієнти даних Best Global Brands

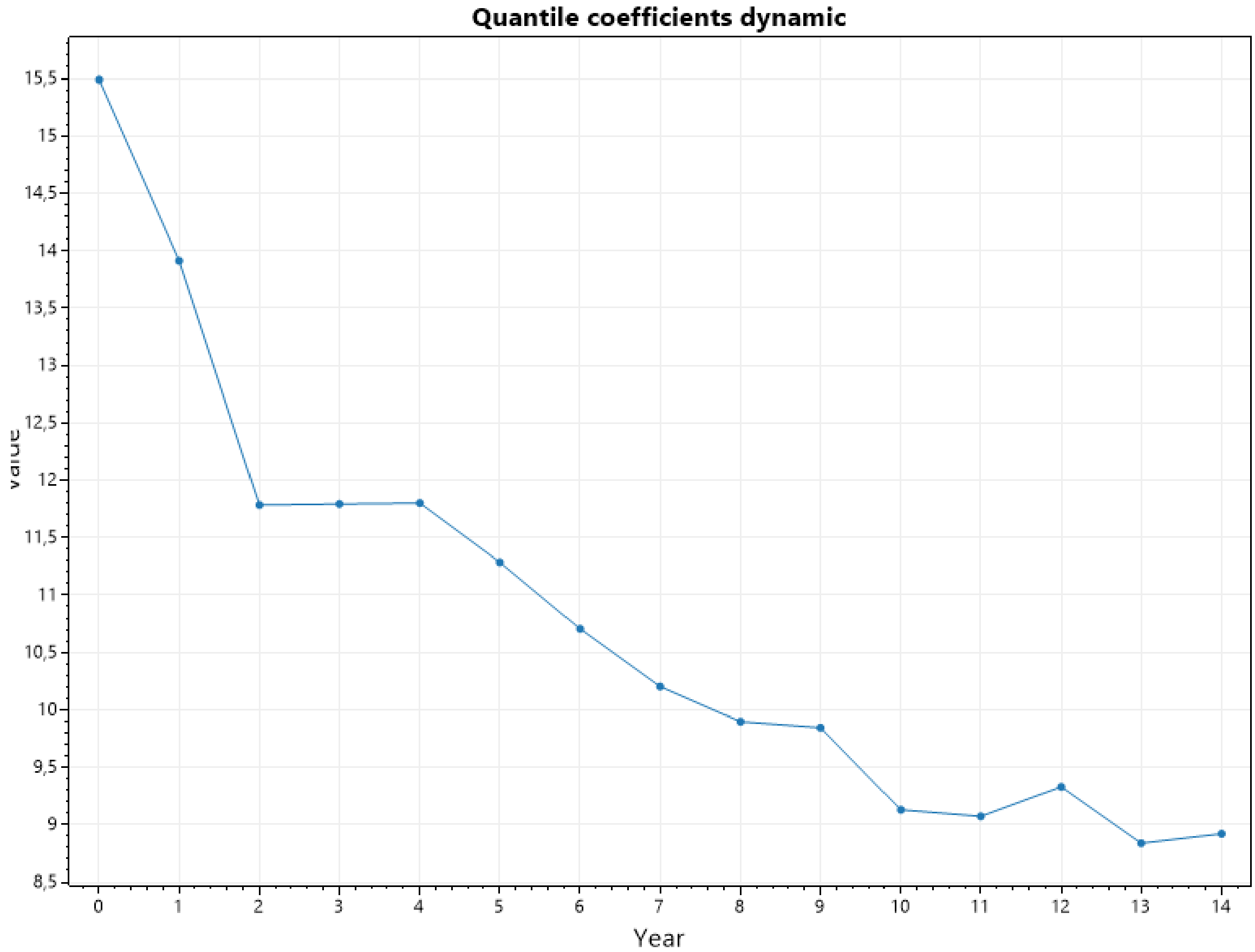

Рисунок 3.7 Квантильні коефіцієнти даних Best Global Brands

Як бачимо на рисунках, значення цих коефіцієнтів зростає майже щороку. Це може казати або про збільшення вартості брендів на перших позиція, або про зменшення вартості брендів на останніх позиціях. При розрахунку децильних коефіцієнтів отримали числа більші ніж при розрахунку квантильних та квартильних коефіцієнтів. Це може казати про те, що чим вище у списку бренд, тим краще він зросте у вартості за рік.

# ВИСНОВКИ

## Розділ 1

Було дано поняття бренду, інформаційної системи, статистичного аналізу та розподілу. Було оглянуто та порівняно сучасні інформаційні системи статистичного аналізу розподілу даних. Було визначено цінність нового рішення в даній сфері. Було деталізовано завдання розробки та вимоги до програми.

## Розділ 2

Було аргументовано вибір мови написання, бібліотек для збору, обробки та подання даних. Було проведено дослідження в результаті якого було визначено, що багато брендів, скоріше за все, розподілені за степеневими законами розподілу. Було показано алгоритм визначення відповідності статистичного розподілу до степеневих законів розподілу. Було дано поняття статистики Колмогорова-Смирнова, методу оцінки максимальної ймовірності для визначення параметру alpha степеневих законів розподілу та метод наближення до параметру xmin степеневих законів розподілу. Було розроблено три програмні модулі, що відповідають за збір й обробку даних і за взаємодію користувача з програмою, на основі оглянутих матеріалів.

## Розділ 3

Було показано роботу програмного продукту. Було досліджено три розподіли вартостей брендів за допомогою створеної системи. Було виявлено що розподіл Best German Brands, ймовірно, розподілений за законом Парето, що розподіл Best Pharma Brands розподілений подібно до закону Ципфа, але не є його проявом, і що розподіл Best Global Brands розподілений за іншим степеневим законом розподілу. Було знайдено децильні, квартильні та квантильні коефіцієнти, та показано їх динамки (де це можливо).

# СПИСОК ДЖЕРЕЛ

# ДЕКЛАРАЦІЯ

Усвідомлюючи свою відповідальність за надання неправдивої інформації, стверджую, що подана кваліфікаційна (бакалаврська) робота на тему: «РОЗРОБКА ІНФОРМАЦІЙНОЇ СИСТЕМИ ДЛЯ СТАТИСТИЧНОГО АНАЛІЗУ РОЗПОДІЛІВ ГЛОБАЛЬНИХ БРЕНДІВ» є написаною мною особисто.

Одночасно заявляю, що ця робота:

- не передавалась іншим особам і подається до захисту вперше;
- не порушує авторських та суміжних прав, закріплених статтями 21-25 Закону України «Про авторське право та суміжні права»;
- не отримувались іншими особами, а також дані та інформація не отримувались у недозволений спосіб.

Я усвідомлюю, що у разі порушення цього порядку моя кваліфікаційна (бакалаврська) робота буде відхилена без права її захисту, або під час захисту за неї буде поставлена оцінка «незадовільно».

16.05.2022